\documentclass[aip,apl,reprint]{revtex4-1}
\usepackage{amsmath}
\usepackage{bm}
\usepackage{dcolumn}
\usepackage{float}
\usepackage{graphicx}
\usepackage{hyperref}
\usepackage{memhfixc}
\usepackage[version=4]{mhchem} 
\usepackage[load=prefixed,load=abbr,alsoload=synchem]{siunitx} 
\usepackage{upgreek} 
\usepackage{xcolor} 

\draft
\begin{document}

\title{Femtosecond-Laser-Induced Spin-Polarized Electron Emission from a GaAs Tip}

\author{Evan Brunkow}
\affiliation{Department of Physics and Astronomy, University of Nebraska-Lincoln, 
Lincoln, Nebraska 68588, USA}

\author{Eric R. Jones}
\email[]{eric.ryan.jones@huskers.unl.edu}
\affiliation{Department of Physics and Astronomy, University of Nebraska-Lincoln, 
Lincoln, Nebraska 68588, USA}

\author{Herman Batelaan}
\affiliation{Department of Physics and Astronomy, University of Nebraska-Lincoln, 
Lincoln, Nebraska 68588, USA}

\author{T. J. Gay}
\affiliation{Department of Physics and Astronomy, University of Nebraska-Lincoln, 
Lincoln, Nebraska 68588, USA}


\date{\today}

\begin{abstract}
It is shown that focusing circularly-polarized \SI{800}{\nm} light pulses of \SI{100}{\fs} duration 
on the tips of \textit{p}-\ce{GaAs} crystalline shards having no negative electron
affinity (NEA) activation 
results in electron emission that is both fast and spin-polarized. 
The \SI{400}{\fs} duration of the emission process was determined by pump/probe 
measurements.
The three samples we investigated produced electron polarizations of 
\SI{13.1\pm0.9}{\percent}, \SI{13.3\pm0.7}{\percent}, and \SI{10.4\pm0.2}{\percent}.
Emission currents ranged between \SI{50}{\pA} and \SI{3}{\nA} with a sample bias of
\SI{-100}{\volt} and average laser power of \SI{100}{\milli\watt}. 
The electron emission exhibited linear dichroism and was obtained under moderate vacuum 
conditions, similar to that of metallic tips.
This source of spin-polarized electron pulses is ``fast'' in the sense that the 
electron emission process is of comparable duration to the laser pulses that initiate 
it. 
\end{abstract}


\maketitle

\section{\label{sec:intro} Introduction}
Sub-picosecond, nanometer-scale, spin-polarized electron sources are currently not 
available.
Such a source is desirable for tests of quantum degeneracy and for ultrafast electron 
microscopy.\cite{Lobastov7069,Lougovski023417,Jones214,Ropers2018}
The first reported observation of free electron antibunching remains controversial, as 
the experimental apparatus could not distinguish between the effects of Coulomb 
pressure and degeneracy pressure.\cite{Kiesel392,Kodama063616,Baym201}
As degeneracy pressure is polarization-dependent, while Coulomb pressure is not, a 
spin-polarized, sub-picosecond, nm-scale source could resolve the controversy.
The best combined spatial and temporal resolution in ultrafast electron microscopes
is provided by nanotip sources triggered by femtosecond laser illumination, as 
photocathodes with a planar geometry are limited in spatial resolution by the 
size of the laser focus.\cite{Zewail187,Bormann173105}
Direct measurements of the electron pulse duration in ultrafast electron microscopy 
have shown that the electron and the illuminating laser 
pulse durations are of the same order.\cite{Hassan425}
Implementing a spin-polarized source into an ultrafast electron microscope would allow
for a novel approach to studying magnetic nanostructures at the fs-scale.\cite{Wan89}

In this work, we present a fast, localized, spin-polarized source of electrons 
obtained from a sharp \textit{p}-\ce{GaAs} bulk [110] crystal shard illuminated with 
femtosecond laser light.
The size of the emission site is approximately \SI{1}{\um} in scale, and the electron 
polarization achieved so far is \SI{13}{\percent}.
The electron emission was studied using methods similar
to those developed to characterize pulsed emission from metallic nanotips.
Such sources are
currently in broad application to produce temporally short electron pulses in beams 
with high brightness.\cite{Barwick142,Hommelhoff077401,Hommelhoff247402,Ropers043907}    
They are referred to as ``fast,'' meaning that the temporal 
response of the emission process is comparable to that of the light pulse duration,
and their spatial resolution has been shown to be determined by the size of the 
emitter and not by the laser focus used.

\begin{figure}[tb]
\includegraphics{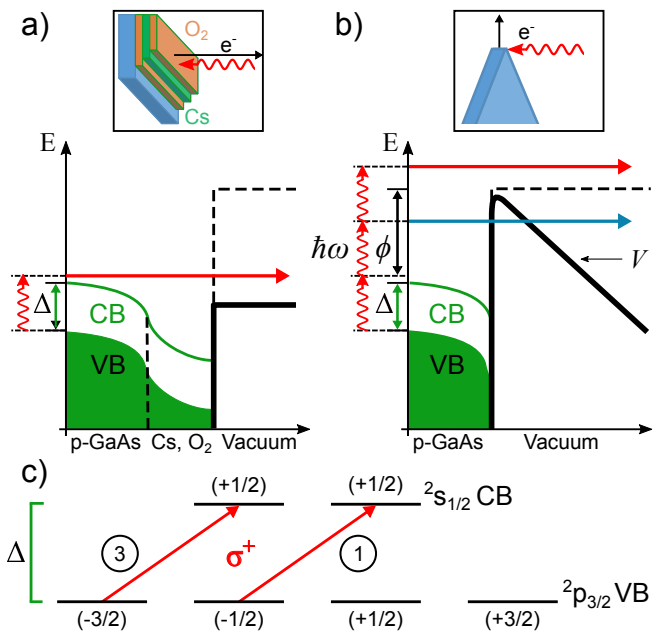}
\caption{\label{fig:Fig1}
\ce{GaAs} energy levels for (a) NEA 
bulk surfaces and (b) a non-NEA shard apex.  
The diagrams indicate bending of both the valence band (VB) and conduction 
band (CB) at the surface due to heavy \textit{p}-doping. 
(a) The vacuum energy (dashed black line), is lowered (solid black line) due to the 
deposition of alternating layers of \ce{Cs} and \ce{O2} (top inset). 
Electron emission from the NEA surface proceeds by the absorption of 
a single photon with energy that exceeds the band gap $\Delta$ of the bulk.   
(b) Multiphoton emission from an uncoated, non-NEA \ce{GaAs} shard apex (see text). 
(c) Allowed transitions at the \ce{GaAs} $\Gamma$-point for absorption of right-hand 
circularly-polarized light by Zeeman ($m_j$) sublevels.  
Selection rules ($\Delta{m}_{j} = +1$) and the relative line strengths (indicated in 
circles) yield a nascent conduction-band electron polarization of $(3-1)/(3+1) = 50\%$ 
for valence-conduction band resonant transitions.\cite{Pierce5484}
}
\end{figure}

Standard CW polarized electron sources use a planar GaAs photocathode that  
must be layered with, e.g., \ce{Cs} and \ce{O2} to lower the vacuum potential below that of the 
conduction band. 
This creates a ``negative electron affinity'' (NEA) condition that allows electron 
emission by absorption of a single photon from a CW laser
(Fig.~\ref{fig:Fig1}(a)).
When circularly-polarized light with an energy near the bandgap~$\Delta$ of \ce{GaAs} 
is used to excite electrons, there is an imbalance in excitation probabilities of 
the two excited ${}^{2}s_{1/2}$ Zeeman substates 
(Fig.~\ref{fig:Fig1}(c)),\cite{Pierce5484} 
causing the emitted electrons to be spin-polarized.
Such sources are used in a variety of fields, including atomic and 
molecular,\cite{Kessler1985,Gay157} 
high-energy,\cite{Prescott347,Androic141803}
and condensed matter physics.\cite{Feder1986,Bergmann046801,Giebels035124}

Alternative planar photocathodes with and without NEA have been developed
to optimize the spin-polarization of the emitted electrons, to 
provide short pulse operation, and to enhance source brightness.
Back-illuminated NEA strained and unstrained thin photocathodes have produced 
\SI{2.5}{\ps} electron pulses with high brightness.\cite{Aulenbacher7536}
There, the electron pulse duration is limited by the slow emission process of 
diffusion through the material. 
A strained \ce{GaAs}-\ce{GaAsP} superlattice with NEA activation resulted in a
\SI{16}{\ps} pulse duration\cite{Kuwahara033102,Kuwahara193101,Kuwahara013108}, 
and was used in a spin-polarized transmission electron microscope. 
The source was determined to have a degeneracy 2 orders of magnitude lower than the
cathode tip used to first study free electron degeneracy,\cite{Kiesel392}
with a source size that was limited by the diffraction limit of the laser focus. 
A planar \ce{GaAs} photocathode with a Ag overlayer a few nm thick has 
functioned as a polarized electron source without NEA activation by utilizing a 
multiphoton electron emission process.\cite{Klaer214425} 
Electron yields were increased by employing local field enhancement through 
plasmonic coupling on the surface of a \textit{p}-doped \ce{GaAs} wafer, 
while the spin-polarization of emitted electrons was largely maintained.
Pulsed $\sim$\SI{100}{\fs} laser light produced a spin-polarization as high as 
\SI{21}{\percent}, with a value of $\sim$\SI{15}{\percent} for illumination at a 
central wavelength of \SI{800}{\nm}.

Tips of magnetized iron and cobalt-coated tungsten have 
been used to produce spin-polarized electrons, although these sources have 
used only CW lasers to date.\cite{Irisawa113031,Niu1055}
Such magnetized sources have a further limitation in that their spin polarization is 
not optically reversible, unlike that of of \ce{GaAs} photocathodes.
An array of etched \ce{GaAs} tips, illuminated with CW laser light for both positive 
electron affinity (PEA) and NEA surface conditions resulted in a maximum polarization 
of \SI{37}{\percent}, but the electrons were not pulsed.\cite{Kuwahara6245}
Implementation of a tip geometry results in field enhancement at the tip apex, which 
increases the yield of emitted electrons.
While a more robust activation surface of layered Cs and Te has been 
demonstrated,\cite{Bae154101}
a tip geometry, as well as a multiphoton emission process, eliminates the need for 
NEA activation that is sensitive to vacuum conditions.\cite{Pirbhai060701}  

The work reported here focuses on obtaining fast, spin-polarized electrons from a sharp 
\textit{p}-\ce{GaAs} bulk [110] crystal shard, which naturally incorporates 
optical reversibility. 
To do this, Ti:Sapph pulsed lasers with wavelengths centered around \SI{800}{\nm},
the appropriate wavelength for single-photon excitation across the band gap, were used to 
induce multiphoton emission without requiring that the samples have NEA.
Fig.~\ref{fig:Fig1}(b) illustrates this.
The vacuum potential (dashed black line) is modified at the surface by the application 
of a negative DC bias voltage~$V$ and the local laser field (solid black line).  
A single photon with energy just exceeding the bandgap~$\Delta$ can promote 
an electron from the valence band to the conduction band. 
Absorption of a second photon can in principle result in emission via tunneling 
through the vacuum potential (blue arrow). 
Absorption of one or more additional photons provides sufficient energy for the 
electron to exceed the additional ionization energy~$\phi$ and escape into the vacuum 
(red arrow).
The \SI{800}{\nm} central wavelength of our lasers accesses the relative 
excitation probabilities for circularly-polarized light that make standard NEA 
\ce{GaAs} sources produce polarized electrons (Fig.~\ref{fig:Fig1}(c)). 

\section{\label{sec:exp} Experiment}
\begin{figure}[tb]
\includegraphics{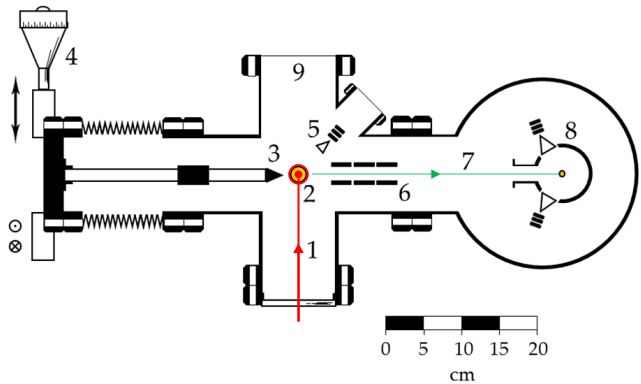}
\caption{\label{fig:Fig2}
The experimental setup for polarimetry and dichroism measurements. 
The pulsed laser beam (1) enters the chamber and hits the off-axis parabolic mirror 
(2) which focuses the laser onto the sample (3).
Note that the beam is propagating out of the plane at (2), indicated by the red circle. 
The sample is mounted on an XYZ translator (4) that allows the sample tip to be 
positioned in the laser focus. 
A CEM (5) can be used to monitor electron emission. 
Transport optics (6) guide emitted electrons (7) 
toward the Mott polarimeter (8) in the adjoining chamber with top (T) and bottom (B)
CEM detectors. 
A \SI[per-mode=symbol]{260}{\litre\per\s} turbomolecular pump (9) evacuates the 
chamber.
}
\end{figure}

We used two apparatuses, the first to measure electron polarization and emission 
dichroism, and the second to study the emission process duration and the emission position 
dependence.
Our first optical setup consisted of a Ti:Sapph oscillator (Griffin, KMLabs) with an 
output that passed through a collimating lens and a periscope placed prior to 
polarizing optics. 
A half-wave plate (HWP) followed by a linear polarizer was used to vary the laser power 
without changing the direction of its linear polarization. 
The beam then passed through a quarter-wave plate to switch its polarization 
from linear to left- or right-handed circular.
A final HWP was used to rotate the plane of polarization of 
linearly-polarized beams. 
The laser then entered the polarization/dichroism vacuum system through a window (Fig.~\ref{fig:Fig2}).
Just before entering the chamber, the width of the laser pulses was measured 
to be \SI{75}{\fs} with a Swamp Optics Frequency-Resolved Optical Gate (FROG).

The vacuum system, with a nominal base pressure of 
\SI{e-7}{\torr}, comprised two sections.
A sample chamber contained an off-axis front-surface Au parabolic mirror to change the 
direction and focusing of the laser to a \SI{20}{\um}-FWHM spot size.  
The \ce{GaAs} shard was mounted on a 3-axis stage to position it in the laser focus. 
A channel electron multiplier (CEM) near the sample monitored 
the electron emission current.  
We also measured the total emission current from the electrically-isolated sample. 
Emitted electrons were directed to a compact, cylindrical Mott 
polarimeter,\cite{Clayburn053302}
comprising two concentric cylindrical electrodes and two CEMs placed symmetrically 
about the entrance that defined the electron scattering plane. 
The central gold-plated electrode was biased at \SI[retain-explicit-plus]{+20}{\kV}, 
whereas the outer electrode and the mouths of the CEMs were biased at 
\SI[retain-explicit-plus]{+500}{\V}.  

To measure the electron polarization, $P_e$, the count rates measured by the top and 
bottom CEMs ($C_T$ and $C_B$) were monitored for electrons produced by light 
pulses that were right-hand circularly-polarized, and then compared with the rates when 
the light helicity was flipped.
The electron polarization, $P_e$, is given as $P_e = S_{\mathrm{eff}}/{A}$,
where
\begin{equation}
\label{eq:eq2}
A = \frac{\chi{} - 1}{\chi{} + 1} ~\mathrm{and}~ \chi = \sqrt{\frac{C_{T}C_{B}'}{C_{T}'C_{B}}}.
\end{equation}
Here, $S_{\mathrm{eff}}$, the ``effective Sherman function,'' is the polarimeter's 
analyzing power, and the primes indicate the CEM rates for left-handed incident laser 
light. 
The advantage of measuring $P_e$ this way is that it eliminates first-order 
instrumental asymmetries.\cite{Kessler1985,Gay157}

Measurements of the linear and circular emission dichroism were 
made to better understand the emission process.
The dichroism, calculated using total emission as measured by the CEM proximate to
the sample, is  
\begin{equation}
\label{eq:eq4}
D \equiv \frac{R_1 - R_2}{R_1 + R_2};
\end{equation}
$R_{1,2}$ is the rate of emission for orthogonal polarizations.

Electron emission from the samples was optimized at the edge of the crystal shard. 
Sharp tip-like shards were made by shattering 
crystalline wafers and using an optical microscope to determine the 
``sharpest'' pieces.\cite{Prins8105} 
When using these, total emission currents between \SI{50}{\pA} and \SI{3}{\nA} 
were obtained with an average laser power of $\sim$\SI{100}{\milli\watt} and a DC 
sample bias of \SI{-100}{\volt}.
 
A second, similar apparatus was used to study emission rates
as a function of the shard apex morphology, to measure the dependence of emission rate 
on laser intensity, and to assess the temporal width of the emission 
process.\cite{Jones263104}
Pulses from a Ti:Sapph oscillator (Spectra Physics Tsunami) 
were focused to a FWHM of \SI{3.6}{\um}.
The laser pulse intensity FWHM, $\tau_{laser}$, was measured to be \SI{100}{\fs}.
The laser power delivered to the shard apex was controlled by a Brewster window 
variable attenuator.
Pulsed electron emission was detected by a microchannel plate (MCP) placed close to
the shard apex, or by an electrometer connected directly to the sample.
Prior to entering the chamber, the primary beam was split into pump and probe 
components in a balanced Mach-Zehnder interferometer.
The delay $\tau$ between pump and probe pulses could be adjusted for values between 
$\pm$\SI{4}{\ps}.

When two temporally-separated light pulses hit the sample, the 
integrated electron emission can be categorized as either ``additive''
or ``super-additive.''  
Additive emission means that the integrated signal is the same as the 
sum of the emission from each pulse individually. 
Super-additivity occurs when the emission is greater than the sum of that due 
to the individual beams.  
Additive emission for $\tau > \tau_0$ shows that the emission process 
does not exceed $\tau_0$; if $\tau_0 \approx \tau_{laser}$, the emission process is 
``fast'' as defined in the Introduction. 
Superadditivity for $\tau \gg \tau_{laser}$ implies the process is slow, e.g., due to 
thermally assisted 
processes.\cite{Barwick142,Hommelhoff077401,Kealhofer035045,Jones263104}

\begin{figure*}[t]
\includegraphics{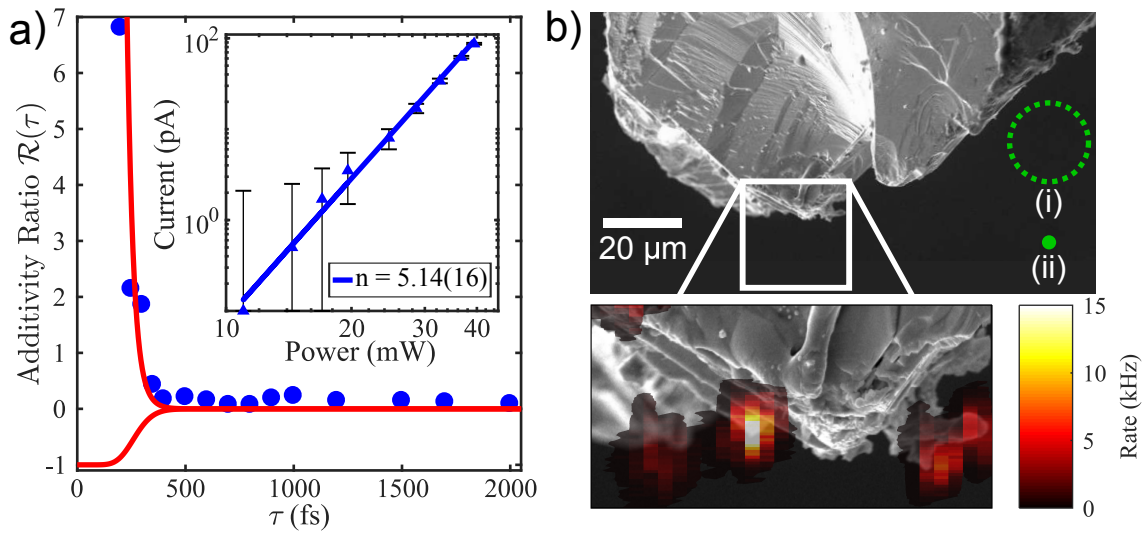}
\caption{\label{fig:Fig3}
Emission data from a \ce{GaAs} shard ``tip.'' 
In (a), the additivity ratio $\mathcal{R}$ is plotted as a function of $\tau$ (blue 
circles).
The red line is the theoretical curve obtained with an electric field width of 
\SI{160}{\fs} and an $I^5$ intensity dependence.
The bifurcation of the $\mathcal{R}(\tau)$ curve for 
$\tau <~ $\SI{400}{\fs} is due to the flipping of the sign in Eq.~\ref{eq:eqmodel2}
of $E_{probe}$, and corresponds to the envelope function for the rapidly oscillating 
autocorrelation interference pattern in this region. 
The power dependence of emission is plotted in the inset.
(b) A scanning electron microscope (SEM) micrograph of the apex area with an expanded 
square section \SI{20}{\um} on a side. 
The laser focal spot size from the polarization measurements (dashed green circle, (i)) is 
shown to the scale of the top micrograph and compared to the focal 
spot size (solid green circle, (ii)) used for the measurements shown in (a).
Localized emission from the shard's sharpest features (inset) indicates that multiple 
sites may have been emitting in the polarization and dichroism measurements.
}
\end{figure*}

\section{\label{sec:results} Results}
We first consider the electron pulse emission process. 
Electron emission from nanotips, if measured to be both
nonlinear and additive for $\tau > \tau_{laser}$, has been shown to be fast.\cite{Barwick142,Hommelhoff077401,Hommelhoff247402} 
Our electron emission current shows non-linearity as a function of intensity. 
It fits with a power law of $n = 5.14(16)$ (Fig.~\ref{fig:Fig3}(a) inset; blue line).
The Keldysh parameter for a solid, $\gamma$, characterizes the emission.
For $\gamma \gg{1}$, field emission is dominated by multiphoton 
processes.\cite{Bunkin896,Kruger074006} 
Given our focal spot sizes of \SI{20}{\um} and \SI{3.6}{\um}, and an average power 
that never exceeded \SI{150}{\milli\watt}, our Keldysh parameter readily satisfied this
condition in all our experiments and supports our simple multi-photon model.
The fifth order non-linearity indicates a five-photon process.
(This result is in excess of the three-photon process illustrated in 
Fig.~\ref{fig:Fig1}(b)).
Generally speaking, the order of the multiphoton process in a given
sample can vary with the details of the emitting surface, its local surface electric 
field, and the nature of surface states near the emission point.
\cite{*[{See, e.g., Section 3.1 of }] [{}] Schmidt223}

Pump-probe measurements as described above were used to determine if the emission was 
additive.\cite{Barwick142,Jones263104} 
The additivity ratio is defined as 
\begin{equation}
\label{eq:eqmodel1}
\mathcal{R}(\tau) \equiv \frac{R_{both}(\tau) - \left(R_{pump}(\tau) + R_{probe}(\tau)\right)}{R_{pump}(\tau) + R_{probe}(\tau)},
\end{equation}
where $R_{pump}(\tau)$ and $R_{probe}(\tau)$ are the emission rates from the pump and 
probe beams separately at each delay, and the rate $R_{both}(\tau)$ was modeled as  
\begin{equation}
\label{eq:eqmodel2}
R_{both}(\tau) = \int\displaylimits_{-\infty}^{\infty} \left[
E_{pump}(t) \pm E_{probe}(t + \tau)
\right]^{2n} \, \mathrm{d}t.
\end{equation}
The individual pump and probe field amplitudes were modeled as Gaussians with 
$E(t)=E_0 \exp [ - ({t /\tau_{pulse}} )^2 ]$.
The best fit to the data (red line in Fig.~\ref{fig:Fig3}(a)) is obtained 
for $\tau_{pulse} =$~\SI{160}{\fs} ($n = 5$).        
The electron emission process is additive 
($\mathcal{R}(\tau) = 0$ for $\tau >~ $\SI{400}{\fs}) and is thus shown to be 
faster than this value.
Note that this is not a direct measurement of the electron pulse duration. 
Nevertheless, fast emission processes have so far indicated short electron 
pulses.\cite{Kruger074006}

We now turn our attention to electron polarization. 
Measurements of $P_e$ were taken with a \SI{20}{\um}-diameter focal spot
for two focal positions on the three samples we studied. 
In the first ``tip'' position, the focal spot was centered on the shard apex. 
In the second ``shank'' position, the focus center was moved about \SI{15}{\um} away 
from the tip towards the bulk. 
The results of all measurements of $P_e$ and emission dichroism, taken with the 
\SI{20}{\um} focus, are given in Table~\ref{tab:table1}.
\begin{table}
\caption{\label{tab:table1} 
Polarization and dichroism results for circularly- and linearly-polarized light 
incident on either the apex (``tip'') or the bulk (``shank'') of three different shard 
samples. 
}
\begin{ruledtabular}
\begin{tabular}{llll}
\multicolumn{1}{c}{\textrm{Target}} & \multicolumn{1}{c}{\textrm{Light}}        & \multicolumn{1}{c}{$P_e(\%)$}  & \multicolumn{1}{c}{\textrm{D(\%)}}\\
                                    & \multicolumn{1}{c}{\textrm{Polarization}} &                                       &                                   \\
\hline 
\noalign{\vskip 2mm}
\textrm{\#1 Tip}   & \textrm{Circular} & 13.1(9)  & {}       \\
\textrm{\#2 Tip}   & \textrm{Circular} & 13.3(7)  & 4.7(6)   \\
{}                 & \textrm{Linear}   & 0.1(5)   & 41.3(1.0)\\
\textrm{\#3 Tip}   & \textrm{Circular} & 10.4(2)  & 1.8(2)   \\
{}                 & \textrm{Linear}   & 2.6(2.5) & 18.5(6)  \\
\textrm{\#1 Shank} & \textrm{Circular} & 1.7(8.0) & 6.4(1.4) \\
{}                 & \textrm{Linear}   & 1.0(2.1) & 23.7(5)  \\
\textrm{\#2 Shank} & \textrm{Circular} & 3.4(1.6) & {}       \\  
{}                 & \textrm{Linear}   & 5.2(1.0) & {}       \\
\end{tabular}
\end{ruledtabular}
\end{table}
In the ``tip'' position, with circularly-polarized laser illumination, $P_e$ was 
\SI{13}{\percent} for samples 1 and 2, and \SI{10}{\percent} for 
sample 3. 
Note that these results are comparable to those of Ref.~\onlinecite{Klaer214425}.
Variations in the local structure or \textit{p}-doping could be responsible for the differences in 
$P_e$.\cite{*[{See, e.g., Section 3.1 of }] [{}] Schmidt223} 
As expected, when the laser was linearly-polarized, the values of $P_e$ were consistent with zero. 
One exception, which we have yet to understand, was observed with sample 2 in the shank 
position.
We note though that this value of $P_e$ is less than half that of the polarization
measured at the tip with circularly-polarized light. 

Finally, we consider the sample morphology. 
The electron emission rate was found to depend sensitively on the position of the 
laser focus at the sample. 
Fig.~\ref{fig:Fig3}(b) shows a plot of the emission rate measured in a \SI{20}{\um} 
square area of a shard apex.
The two laser focal spot sizes used in this work are shown relative to the size of 
the \SI{20}{\um} scale bar in the top micrograph.
The brightest emission feature was used to measure the emission dependence on 
intensity (Fig.~\ref{fig:Fig3}(a) inset) and to perform the pump/probe measurements. 

Non-zero linear emission dichroism (Eq.~\ref{eq:eq4}) was observed for the GaAs shards
similar to a field emission tip (FET). 
That is, emission is higher when the light's linear polarization is parallel to 
the axis of the tip.\cite{Barwick142,Hommelhoff077401} 
In contrast, emission dichroism is absent for 
standard planar \ce{GaAs} sources.\cite{Kessler1985,Gay157,Pierce5484} 
Dichroism measurements were taken at both focal positions as well. 
At the tip of the \ce{GaAs}, the circular dichroism is small 
($<$\SI{5}{\percent}) and the linear dichroism for tips 1 and 2 are 
\SI{41}{\percent} and \SI{19}{\percent}, respectively.  
Linear dichroism measured for tip 1 drops to \SI{24}{\percent} at the shank, 
possibly because there is less of a tip-like structure with which the light 
interacts. Thus our shard ``tips'' have emission characteristics similar to those of 
FETs\cite{Barwick142,Hommelhoff077401,Hommelhoff247402,Ropers043907} in terms of nonlinearity, additivity, polarization and local morphology 
although it is apparent from Fig.~\ref{fig:Fig3}(b) that the overall shard 
morphology is complex.

In summary, we have demonstrated a source that is able to produce fast pulses of 
polarized electrons from a micrometer-size area. 
This can, in principle, enable the imaging of a small electron spot on a target to measure 
spin-dependent effects with fs-scale resolution. 
The reduced vacuum requirements of this source when compared with NEA \ce{GaAs} 
sources make it easier and less costly to operate. 
Although the observed electron polarization is modest, our results demonstrate  
that this source follows the selection rules illustrated in Fig.~\ref{fig:Fig1}(c). 
Polarization might be increased by having a sharper, more well-defined 
\ce{GaAs} tip, or varying the laser wavelength. 
The parameter space is large and open to future study.
Through the use of chemical etching and ion milling, it is possible to shape the tip. 
An optical parametric amplifier can be used to explore the wavelength-dependence of 
polarization. 
Investigation of the effects these parameters have on the total yield and polarization of the 
emitted electrons is needed. 

\begin{acknowledgments}
We thank M. Becker for useful conversations and S. Keramati for taking the electron
micrographs shown in Fig.~\ref{fig:Fig3}(a). 
This work has been funded by NSF awards PHY-1206067 and 1505794 (TJG),
EPS-1430519 (HB and TJG), and PHY-1602755 (HB).
\end{acknowledgments}

\providecommand{\noopsort}[1]{}\providecommand{\singleletter}[1]{#1}%

\end{document}